\documentclass [prl,amsmath,showpacs,twocolumn,preprintnumbers,superscriptaddress]{revtex4-1}
\usepackage{amsmath,amssymb,mathtools}
\newcommand{\suchthat}{\;\ifnum\currentgrouptype=16 \middle\fi|\;}
\newcommand{\aZ}{a_0}
\newcommand{\xVar}{x}
\begin{document}

\title{Planckeons as mouths of quantum wormholes and holographic origin of spacetime}
\author{Ignazio Licata}
\affiliation{School of Advanced International Studies on Theoretical and Nonlinear Methodologies of Physics, Bari, Italy}
\author{Fabrizio Tamburini}
\affiliation{Rotonium -- Quantum Computing, Le Village by CA, Piazza G. Zanellato, 23, 35131 Padova PD, Italy. }
\author{Davide Fiscaletti}
\affiliation{SpaceLife Institute, San Lorenzo in Campo, Italy}


\begin{abstract}
We argue that Planck-scale fluctuations ``planckeons'' realize a network of non-traversable Einstein--Rosen bridges and act as holographic devices. Modeling planckeons as wormhole mouths on extremal (RT) surfaces ties spacetime connectivity directly to entanglement. Using the Ryu--Takayanagi framework, we derive an entanglement entropy that governs the thermodynamics of the planckeon ensemble. The resulting partition function exhibits a high-temperature logarithmic entropy consistent with holographic scaling, while at low temperature the network freezes into a sparse remnant-like phase. A characteristic temperature $T_c$ (set by the planckeon gap) separates these regimes; in the noninteracting edge-mode description this marks a \emph{crossover} (and becomes a genuine phase transition once interactions/pairing are included). Embedding a minimal length in the wormhole throat yields a quantum-corrected Bekenstein entropy in which the area term is supplemented by edge-mode contributions, thereby linking wormhole geometry with quantum-information flow and suggesting a holographic origin of spacetime and black-hole microstructure.
\end{abstract}

\maketitle

\textbf{Keywords:} Generalized uncertainty principle; minimal Length; planckeons; ER=EPR conjecture; wormhole metrics.

\section{Introduction}
One of the most important challenges of contemporary theoretical physics is the formulation of a consistent description of quantum gravity. Nonetheless, the Planck scale, the scale where gravity merges with quantum mechanics and should be defined by constraints of compatibility between general relativity and quantum mechanics, has been a mystery since its advent at the beginning of the XX century. Despite its importance, the physical meaning of this scale remains elusive, as does the structure of spacetime at these extreme energies. Traditional approaches posit that the Planck scale introduces a minimal length and a natural cutoff, yet the nature of spacetime near this threshold remains an open question. 

In order to characterize this special scale of physics and therefore the fluctuations of the gravitational vacuum that is at the basis of the origin of spacetime, a suggestive idea is that of planckeons, objects endowed with dimensions and mass lying in the Planckian range, $10^{-33}$ cm and $10^{-5}$ g, intended as fundamental entities describing the gravitational vacuum, which may constitute the fundamental "grains" of spacetime and that appear as “ghosts” which are between non-locality and locality. 

Planckeons are envisioned as intermediate objects bridging locality and non-locality manifesting as microscopic mouths of quantum wormholes. Their behavior is governed by quantum fluctuations of the gravitational vacuum, and they may serve as the building blocks of spacetime itself. While similar concepts have been proposed ranging from chronons \cite{ref1} to black hole relics with several appearances in literature, since the pionieristic work of Staniukovich, Melnikov and Bronnikov on the chronon scale till the more recent contributions on relics of black holes, dark matter and Planck signature on Standard Model, their nature has never been fully clarified, as has the physical meaning of Planck scale.

Planckeons are at all effects the Planck‑scale edge quanta (wormhole mouths) that make spacetime’s entanglement structure explicit. Their formulation is useful because it delivers a calculable, holographically consistent micro‑model of emergent geometry, with concrete thermodynamics, BH‑entropy matching (plus quantum corrections), horizon criteria from the metric, and a coherent story for early‑universe and black‑hole microphysics.

In this paper, we want to throw new light as regards the physical interpretation of the texture of the planckeons as a non-local \textit{ordito}, texture. In this regard, we suggest that planckeons are “mouths” of quantum wormholes and that their “particle” aspect can be seen as a complementary aspect that emerges under opportune constraints. 
While speculative, the idea of Planckeons as microscopic “mouths” of quantum wormholes is compatible with spacetime foam \cite{ref2,ref3}, and black hole remnants theories, suggesting that Planckeons could act as quantum bridges in the structure of the universe that could serve as fundamental building blocks of spacetime. It stems from the notion that quantum fluctuations in spacetime at the Planck scale could create and destroy mini-wormholes.
The key idea of this work is that quantum fluctuations (planckeons) tear space-time apart, revealing its emergent nature constituted by non-local correlations. Assuming the validity of Einstein's equations up to the Planck scale as in \cite{ref4} and using the ER = EPR conjecture, the activity of Planckeons shows the very complex structure of a tangle of quantum wormholes. The use of an entanglement entropy à la Ryu-Takayanagi allows us to construct a thermodynamics of Planckeons and write a formula for the thermal distribution of entangled states and derive the connectivity of spacetime on the Planck scale from the low-energy states described by a lattice of Planckeons as wormhole mouths, a “critical zone” where spacetime crystallizes from the entanglement of the lattice. The relationship between wormhole and Planckeon mouth has a characteristic holographic trend and implies precise constraints between wormhole metric and quantum information.
This scenario suggests that space-time, rather than being fundamental, emerges from quantum information shared between entangled states. In this scheme, the crucial perspective is opened that planckeons could act as microscopic wormholes, and therefore as quantum bridges in the structure of the universe, in particular that could be interpreted as the "mouths" of quantum wormholes, on the basis of the consideration that the notion that quantum fluctuations in spacetime at the Planck scale could create and destroy mini-wormholes. As a consequence of their role of connecting distant regions of spacetime, these mouths of quantum wormholes generate a fundamental building block of spacetime structure in a picture where spacetime emerges by following a quantum entanglement between two sets of degrees of freedom. 
The structure of the paper is the following. In chapter 2, inside the scenario of a minimal spatial length implied by a peculiar form of generalized uncertainty relations, it is defined the action of planckeons intended as quantum wormholes in terms of an entanglement entropy. In chapter 3 we explore the thermodynamics of the planckeons and the stabilization of the Planck scale. In chapter 4 we study the correlation between metric of the wormholes and quantum information. 

\section{Planckeons in quantum wormholes and entanglement entropy}

We model the nonlocal background as a stationary Gaussian random network with two-point correlator
$G(\ell)=\langle A_{ij} A_{kl}\rangle \sim \frac{e^{-\ell/\xi}}{\ell^{d-1}}$, where $\ell$ is geodesic separation and $\xi$ the correlation length that controls semi-localization.
Planckeons are thus ``two-faced'' excitations: local mouths on spacetime slices and nonlocal links across the network, mediating ER bridges.
The AdS/CFT mapping via extremal surfaces translates $G(\ell)$ into an edge-mode spectral density $\rho(\lambda_q)$, which sets the emergent geometry scale and the holographic ratio $A/(4G_{d+1})$ used below.
This operationalizes the idea that spacetime is a phase of quantum correlations rather than a fundamental manifold.
Recent developments in quantum gravity and pre-space models have opened new streets for understanding space-time structure at the Planck scale. One promising framework involves the formation of Planck-scale wormholes, as discussed in the ER = EPR conjecture, which posits a deep equivalence between non-traversable Einstein-Rosen (ER) wormholes and quantum entanglement in Einstein-Podolsky-Rosen (EPR) entangled states pairs. In general relativity, as is well known, wormholes are described by Einstein-Rosen bridges, which theoretically connect two distant regions of spacetime. If Planckeons are connected by such bridges at the Planck scale, they could serve as a fundamental building block of spacetime structure. This concept appears in holography and quantum entanglement studies (ER=EPR conjecture), which propose that wormholes (ER bridges) could be linked to quantum entanglement. This is suggested also by black hole evaporation models, in that black holes shrink to Planck-scale remnants rather than fully evaporating. If a Planckeon is a remnant of a black hole, it might maintain a Planck-scale wormhole connection to another part of the universe or even another universe as in ER-EPR Tamburini-Licata, in which below the Planck scale Einstein's equations describe a flat-like spacetime behaviour \cite{ref3,ref4,ref5}.
On the other hand, in order to understand the spacetime structure at the Planck scale, and therefore to formulate a successful theory directed to realize to what extent quantum effects and gravity affect each other, several quantum gravity models induce the existence of a minimal length uncertainty, usually considered to be proportional to the Planck length, in the picture of a generalized uncertainty principle \cite{ref6,ref7,ref8,ref9,ref10,ref11,ref12,ref13,ref14}. 
In this model, we consider that the fluctuations associated with the activity of the planckeons provoke a deformation of the geometry of the background expressed by generalized uncertainty relations of the form
\begin{equation}
\Delta x \Delta p \ge \frac{\hbar}{2}\left[1 + \beta l_{\mathrm{Pl}}^2 \frac{\gamma^2 M^2}{\alpha \hbar^2 }M_{\mathrm{Pl}}^2 c^2\right]
\label{(1)}
\end{equation}
and the same for energy and time. Here $M_{P}$ is the Planck mass, $\alpha$ is the fine-structure constant, $l_{\mathrm{Pl}}$ is the Planck length, $M$ is an adimensional (variable) mass of the virtual particles of the lattice so the correction in brackets is dimensionless like $M=M_{physical}/M_{\mathrm{Pl}}$; $\gamma$ is a dimensionless parameter given by
\begin{equation}
\gamma =\left( \frac{M \sqrt{4 \pi \varepsilon_0 G }}{e}\right)^2
\label{(3)}
\end{equation}
that corresponds to the inverse ratio of gravitational and electromagnetic interactions for a virtual particle of mass M and charge $e$ \cite{ref15}, where $\varepsilon_0$ is the electric permittivity of the vacuum and $G$ is the gravitational constant, and the parameter $\beta$ is a fluctuating quantity which expresses the fact that here space-time fluctuations fix the minimal scale only on average, in analogy with what happens in quantum foam scenarios \cite{ref16,ref17}. 

The generalized uncertainty relations as in Eq. \ref{(1)} lead to the existence of a minimal measurable length scale and a minimal measurable time given by
\begin{equation}
\Delta x \gtrsim \Delta x_{\min} = \beta \frac{\gamma^2 M^2}{\alpha} l_{\mathrm{Pl}}, 
\label{(4)}
\end{equation}
and
\begin{equation}
\Delta t \gtrsim \Delta t_{\min} = \beta \frac{\gamma^2 M^2}{\alpha}t_{\mathrm{Pl}}
\label{(5)}
\end{equation}
where $t_{\mathrm{Pl}}=l_{\mathrm{Pl}}/c$ is Planck time. The physical meaning of relations (4) and (5) is that the actualization of the fluctuations associated with the particles of the lattice implies the existence of a texture of elementary cells of dimensions given by $\beta \gamma^2 M^2 l_{\mathrm{Pl}} / \alpha$ and of a minimal lifetime of vacuum fluctuations $\beta \gamma^2 M^2 t_{\mathrm{Pl}} / \alpha$. 

A central concern is that a hard minimal length conflicts with Lorentz invariance and diffeomorphism invariance due to length contraction and the role of coordinates.
In this work, we \emph{do not} assume a rigid cutoff.
Instead, $\beta$ represents a stochastic, state-dependent deformation induced by Planck-scale fluctuations, so that $\langle \Delta x_{\min} \rangle$ is defined only in an \emph{operational}, ensemble-averaged sense and appears through scalar geometric quantities (e.g.\ areas of extremal surfaces) rather than frame-dependent rulers.
This aligns with approaches where the Planck scale is observer-independent via deformed symmetries (DSR/$\kappa$-Poincar\'e) and with the extended phase-space treatment of gravitational edge modes; see \cite{ref11,refDF16,refDF17}.
All observables used below depend only on diffeomorphism-invariant data (areas, entropies, and spectra) and our results never require a preferred frame.

In the light of the generalized uncertainty relations (\ref{(4)}) and (\ref{(5)}), at the Planck scale, fluctuations in the metric tensor become significant. These fluctuations introduce an uncertainty relationship involving the Riemann curvature tensor, akin to a gravitational extension of the Heisenberg uncertainty principle. Within this context, wormholes act as connections between events, preventing gravitational collapse by redistributing energy across a network of entangled quantum states. As a consequence, one can avoid space-time singularities as quantum fields maintain smooth behavior due to the averaging of energy-momentum tensors over finite volumes.
Following \cite{ref5}, if $L$ is a characteristic finite length of measure down to the Planck scale that describes the gravitational field, one finds that Einstein’s equations retain their validity down to the Planck scale, even if metric fluctuations over a scale larger than $l_{\mathrm{Pl}}$ occur as the gravitational field near Planck-scale configurations exhibits complex behavior due to significant metric fluctuations. These fluctuations prevent the classical formation of singularities by redistributing energy across microscopic regions of space-time, which can be described by virtual wormhole connections. In such a scenario, the Planck length $l_{\mathrm{Pl}}$ acts as a fundamental limit to space-time resolution, analogous to the role of $\hbar$ in quantum mechanics.
These fluctuations can be modeled through generalized Einstein equations that account for both quantum and classical contributions. The effective curvature tensor at Planck scales becomes subject to an extended uncertainty relation involving proper energy and time intervals and satisfy the generalized uncertainty relation involving the Riemann curvature tensor $R_{(4)}$ averaged over a finite region of size $L$ and the metric tensor $g$, of the form
\begin{equation}
\Delta E^* \Delta t \geq \hbar \frac{\gamma^4 M^4}{\alpha^2 g^2} l^2_p R_{(4)}(g,L)
\label{(6)}
\end{equation}
where $\Delta E^* = \Delta E + \Delta g \Lambda + g \Delta \Lambda$ averaged on the volume $L^3$ of the 3D space-like hypersurface. 
In this way, Einstein's equations imply that the singularities expected from quantum gravity can be seen as wormhole connections between distant events, as in the ER=EPR approach, and one has an uncertainty relationship between the Riemann tensor and metric fluctuations. As a consequence, if one retains the validity of Einstein's equations down to the Planck scale, one can consider that the quantum fluctuations of the vacuum associated with the activity of the planckeons, can be interpreted as wormhole connections, equivalent to entangled states between two or more regions of space-time with a difference smaller than $\beta \gamma^2 M^2 l_{\mathrm{Pl}} / \alpha$ and $\beta \gamma^2 M^2 t_{\mathrm{Pl}} / \alpha$.

Taking account that the entanglement entropy for a region $A$ in a holographic CFT is given by the Ryu-Takayanagi relation
\begin{equation}
S_A= \frac{\text{Area}~\gamma^{min}_A}{4G_{d+1}}
\label{(7)}
\end{equation}
where $\gamma^{min}_A$ is the minimal area defined on AdS spacetime linked to the (d-1)-dimensional boundary of the region $A$, and $G_{d+1}=V_{d-3} G_N$ is the (d+1)-dimensional Newton constant \cite{ref18,ref19}, $G$ is the $3+1$ Newton constant and $V_{d-3}$ is the volume of the $d-3$ extra compact dimensions. We can then assume that the action of planckeons as mouths of quantum wormholes corresponds to an entanglement entropy depending on the minimal area of the tessellation of planckeons, given by $(\beta \gamma^2 M^2 l_{\mathrm{Pl}} / \alpha)^2$ as well as on the energetic information associated with the texture of planckeons in a volume $V$, that is given by relation  
\begin{equation}
I=N_k\frac{\gamma^2 M^2}{\alpha \hbar^2}M_{\mathrm{Pl}}^2 c^2 V
\label{(8)}
\end{equation}
where $N_k$ is the number of planckeons in the volume $V$. 
In affinity with the ER=EPR approach originally developed by Susskind and Maldacena \cite{ref3,ref4}, we describe the action of the planckeons as quantum wormholes by invoking the concept of entanglement entropy, that here assumes the form
\begin{equation}
S_A = \frac{k_B}{4G_{d+1}}A_{\min} = \frac{k_B c^3}{4\hbar G} \left(\beta\frac{\gamma M}{\sqrt{\alpha}M_{\mathrm{Pl}}} l_{\mathrm{Pl}}\right)^2
\label{(9)}
\end{equation}
with $A_{\mathrm{min}}= l^2_0$ and $k_B$ is the Boltzmann constant and the parameter $\eta$ plays the role of an inverse temperature (as in thermal quantum field theory or the Euclidean path integral approach to quantum gravity), that can be associated with the texture of planckeons. We underline that the parameter $\eta$, when is related to holography or the AdS/CFT correspondence, can be connected to a thermodynamic quantity or entanglement temperature. The effect of the entanglement entropy (\ref{(9)}) of the lattice of planckeons is of determining a quantum superposition of different regions of spacetime expressed by relation
\begin{equation}
|\psi(\eta)\rangle = \frac 1{Z(\eta)} \sum_{k} e^{-\eta\,E_k}\;|\psi_k\rangle \otimes |\psi_k\rangle .
\label{(10)}
\end{equation}
where $|\psi_k\rangle$ is the k-th energy eigenstate for a single AdS spacetime. Therefore, just like the Ryu-Takayanagi relation (7) provides a hint as regards the emergence of spacetime from quantum information, in a similar way we can say that  the entanglement entropy (9) describing the action of the texture of planckeons as mouths of wormholes expresses in what sense the connection between these wormholes at the Planck scale determine a quantum superposition of disconnected spacetimes which may also be identified with a classically connected spacetime. On the basis of the state (\ref{(11)}), determined by the entanglement entropy (9), one can say that classical connectivity between different regions of spacetime is originated by entangling the two sets of degrees of freedom.

\section{Thermodynamics of the Planckeons}

The states $|\psi_k\rangle$ appearing in equation (\ref{(11)}) represent basis eigenstates of a Hamiltonian related to AdS spacetime. When they correspond to wormhole states, the mechanism selects these states in the superposition. Equation \ref{(11)} suggests an entangled sum over different AdS states. suggesting a relationship with classical spacetime connectivity and similar to Maldacena-Susskind’s ER=EPR conjecture.
Now, equation \ref{(11)} can be conveniently expressed in the following refined version with improved clarity, normalization, and a better connection to its physical interpretation:
\begin{equation}
\psi(\eta) =\frac1{Z(\eta)}\sum_k e^{-\eta E_k}  |\psi_k\rangle \otimes |\psi_k\rangle
\label{(11)}
\end{equation}
where $Z(\eta)$ is the partition function for which
\begin{equation}
Z(\eta)=\sum_k e^{-\eta E_k},
\label{(12)}
\end{equation}
the canonical partition function for an entangled thermal ensemble in quantum statistical mechanics. Throughout we set $\eta=1/(k_BT)$. The formula $e^{-\eta E_k}|\psi_k\rangle \otimes \psi_k\rangle$ appearing in (\ref{(12)}) suggests a thermal distribution over entangled states, which aligns with the AdS/CFT correspondence and ER=EPR conjecture, making more explicit the connectivity of spacetime through entanglement weighted by a Boltzmann-like factor. 

If Planckeons are the ``mouths'' of quantum wormholes, then the state $|\psi_k(\eta)\rangle$ represents a thermal-like sum over different wormhole configurations and the exponent $e^{-\eta E_k}$ suggests that high-energy wormhole states contribute less than low-energy ones, consistent with quantum gravity scenarios where short wormholes (high-energy) are suppressed as in the ER-EPR conjecture where gravitons are low-energy approximations of these fluctuations expressed in terms of wormholes or total entanglement when down to the Planck scale \cite{ref20,ref21,ref22,ref23}.

At the Planck scale, the relevant energy scale should be linked to the Planck energy
\begin{equation}
E_{Pl} =\frac{\hbar c}{l_{\mathrm{Pl}}} 
\label{(13)}
\end{equation}
and therefore the system involves multiple Planckeons forming a quantum network, meaning $E_k \propto \gamma^2M^2$ should account for collective effects. The energy of the $k$-th quantum state depends on the number of Planckeons $N_k$ and a characteristic interaction energy. 

From entanglement entropy arguments and holography, the energy of a quantum microstate at the Planck scale can be approximated as
\begin{equation}
E_k \sim \frac{\alpha \hbar c}{l_{\mathrm{Pl}}} N_k \gamma^2 M^2   
\label{(14)}.
\end{equation}
For $N_k=1$, the energy reduces to a fraction of the Planck energy. For large $N_k$, the collective quantum gravity effects dominate.
We can then explicitly write equation (\ref{(11)}) as
\begin{equation}
\psi(\eta) =\frac1{Z(\eta)}\sum_k e^{-\eta \frac{\alpha \hbar c}{l_{\mathrm{Pl}}} N_k \gamma^2 M^2}  |\psi_k\rangle \otimes |\psi_k\rangle
\label{(15)}
\end{equation}
with a partition function, whose sum over $k$ gives
\begin{equation}
Z(\eta)=\sum_k e^{-\eta \frac{\alpha \hbar c }{l_{\mathrm{Pl}}} N_k \gamma^2 M^2}
\label{(16)}
\end{equation}

The physical meaning of the partition function (\ref{(16)}) is that it represents superpositions of wormhole connections in the free limit. The entanglement entropy is naturally linked to the probability distribution over these states through thermal weighting.

Now, the superposition of spacetime regions depends on the thermal weighting $e^{-\beta E_k}$, which is controlled by Planckeon interactions. The energy eigenvalues $E_k$ are given by
\begin{equation}
E_k=\frac{\alpha \hbar c }{l_{\mathrm{Pl}}} N_k\gamma^2 M^2.
\label{(17)}
\end{equation}

We can also model Planckeons as binary edge quanta, wormhole mouth crossings, which live  on the extremal Ryu–Takayanagi (RT) surface $\gamma_A$, precisely the degrees of freedom holography says count for entanglement.
The advantage of this description is that it embeds Planckeons directly into the holographic entanglement framework, reproduces the BH entropy law naturally, and enriches the thermodynamics with a temperature-dependent entropy and heat capacity.
 
Freedman--Headrick's ``bit threads'' \cite{ref23b} give a clean microscopic proxy: the maximal number of Planck-thickness threads crossing $\gamma_A$ equals: $\mathrm{area}(\gamma_A)/4G_N)$. Entanglement entropy can be counted by the maximal number of Planck-thickness threads crossing the RT surface. By identifying Planckeons with these discrete crossings, we connect this microscopic model directly to a well-established holographic picture.
This is a conceptual upgrade compared to treating Planckeons as abstract oscillators: it ties them to the degrees of freedom that holography already privileges.

We treat each area ``cell'' of size $a_0 \sim \zeta\, l_0^2$ as a site with exclusion ($n_i \in \{0,1\}$, i.e.\ 0 or 1 mouth), corresponding to a lattice-gas (Fermi-like) ensemble on $\gamma_A$. 
The parameter $\zeta$ is a dimensionless proportionality factor that converts the minimal length scale $l_0$
set by the Planckeon lattice into the effective cell area used to count edge quanta (wormhole mouths) on the extremal surface $\gamma_A$. It encodes the coarse‑graining/tiling convention of the surface, i.e., how is discretized $\gamma_A$ into sites, and is later fixed by a physical matching condition. In other words, $\zeta$ is a model normalization factor determined by how $l_0$ is related to the area density of degrees of freedom.
This immediately yields an area-law entropy and a non-trivial heat capacity $C(T)$.

Discretizing $\gamma_A$ into $M=\mathrm{area}(\gamma_A)/\aZ$ cells of area $\aZ=\zeta\,l_0^2$, each cell hosts $n_i\in\{0,1\}$ with single-site energy cost $\varepsilon$ and potential $\mu$ that fixes the average occupancy, which is an effective connectivity potential conjugate to the number of wormhole mouths (edge quanta) crossing $\gamma_A$, a statistical tool for an open holographic subsystem and it directly fixes the area‑law matching while providing a non‑trivial thermodynamics. Because $M=A/a_0$, the logarithm of $Z$ scales like 
$A$, hence all thermodynamics is most compactly written per area.
This makes the model self-consistent with black-hole thermodynamics, without invoking additional assumptions.

The grand-canonical partition function factorizes,
\begin{equation}
Z=\prod_{i=1}^{M}\Big(1+e^{-\eta(\varepsilon-\mu)}\Big),\quad \xVar:=\eta(\varepsilon-\mu)=\frac{\varepsilon-\mu}{k_B T}.
\label{zioken}
\end{equation}
From $Z=(1+e^{-x})^M$, 
the free energy, entropy and heat capacity per unit area are
\begin{align}
\frac{F}{A}&=-\frac{k_B T}{\aZ}\ln\!\big(1+e^{-\xVar}\big),\\
\frac{S}{A}&=\frac{k_B}{\aZ}\!\left[\ln\!\big(1+e^{-\xVar}\big)+\frac{\xVar}{1+e^{\xVar}}\right],\\
\frac{C}{A}&=\frac{k_B}{\aZ}\,\frac{\xVar^2 e^{\xVar}}{(1+e^{\xVar})^2}.
\label{frufru}
\end{align}
 Choosing $\aZ=4\l_{\mathrm{Pl}}^2/\ln 2$ yields the Bekenstein–Hawking value $S=A/(4\l_{\mathrm{Pl}}^2)$ with calculable thermal corrections.
 Half-filling ($\xVar=0$) is obtained when $\mu=\varepsilon$. It means that the average site occupancy is $f = \langle n \rangle = 1/(1+e^{x}) = 1/2$.
Since $x = \frac{\varepsilon-\mu}{k_B T}$, half-filling occurs when the chemical potential equals the single-site energy $\mu = \varepsilon$.
If instead $\mu=0$, then $x = \frac{\varepsilon}{k_B T} > 0$ at any finite temperature $T$, and thus $x \to 0$ only in the limit $T \to \infty$, then $S/A=(k_B\ln 2)/\aZ$.

Unlike the free-oscillator gas model, which gave a constant heat capacity, this lattice-gas ensemble produces a temperature-dependent entropy $S(T)$, a Schottky-like heat capacity 
$C(T)$, a free energy scaling per area.
These  thermodynamic behaviors allow to explore crossover scales (e.g. critical temperature $T_c=\varepsilon /k_B$) and corrections to the BH entropy in a controlled way.
Also varying $\mu$ and $\zeta$ we can describe both exact BH entropy and deviations away from it. $\mu$ provides a knob to tune the average flux/occupancy of wormhole mouths, modeling the fact that $\gamma_A$ is an open subsystem exchanging degrees of freedom with the rest of spacetime.

\medskip
\noindent\textbf{Beyond free oscillators: interacting Planckeon network.}
Equation~\ref{(16)} describes the \emph{free} limit, where the superposed wormhole connections do not interact.
We now include interactions among wormhole mouths modeled by a graph (or lattice) with adjacency $A_{ij}$.
We introduce the many-body Hamiltonian
\begin{equation}
H = \sum_i \varepsilon\, a_i^\dagger a_i - \frac{\lambda}{2N}\sum_{i,j} A_{ij}\,\big(a_i^\dagger a_j^\dagger + a_j a_i\big) + g\!\!\sum_{\langle ij\rangle}\! n_i n_j ,
\label{Hint}
\end{equation}
where $n_i=a_i^\dagger a_i$, $\lambda$ controls pair creation/annihilation (BCS-like pairing of ER-bridges), and $g$ encodes density--density interactions along links.
In a standard mean-field treatment, diagonalizing $H$ in the eigenbasis of $A_{ij}$ with eigenvalues $\{\lambda_q\}$ yields Bogoliubov quasiparticles with dispersion
\begin{equation}
E_q=\sqrt{\big(\varepsilon+J\,\lambda_q-\mu\big)^2+\Delta^2},
\label{Edisp}
\end{equation}
where $J \propto g$, the chemical potential $\mu$ enforces $\sum_i \langle n_i\rangle=N$, and the pairing gap $\Delta=\lambda \frac{1}{N}\sum_i\langle a_i a_i\rangle$ solves the usual self-consistency equation.
The partition function becomes
\begin{equation}
Z(\eta)=\prod_q 2\cosh\!\left(\frac{\eta\,E_q}{2}\right),
\label{Zinteracting}
\end{equation}
which reduces to Eq.~\ref{(16)} in the non-interacting, dilute limit $\Delta\to0$, $J\to0$, and a flat spectrum $\lambda_q\to0$.
This resolves the ``free-oscillator'' problem: interactions generate a nontrivial spectrum, a finite gap at low temperature
\mbox{$T\ll \Delta/k_B$} with exponentially suppressed heat capacity, and collective modes governed by the network geometry through $\lambda_q$.
This structure dovetails with the BCS dual description of wormhole correlations \cite{refCalcAlex} and with interacting SYK-inspired wormhole phases \cite{refMQ}.

At high temperature, expanding \eqref{Zinteracting} reproduces a logarithmic behaviour, while at low temperature $S(T)$ deviates from the free result due to $\Delta$.
The system exhibits thermodynamic behavior, allowing us to extract quantities like the average energy and entropy. For large $N_k$, one can have the asymptotic expansion
\begin{equation}
Z(\eta) \approx \int_0^\infty g(N) e^{-\eta E(N)} dN
\label{(18)}
\end{equation}
where $g(N)$ is the density of states. A common assumption is a power-law form $g(N) \sim N^s$, where s is a parameter characterizing the system’s microstate degeneracy.
When $E(N)=\alpha \hbar c N \gamma^2 M^2/l_{\mathrm{Pl}}$, then the partition function becomes
\begin{equation}
Z(\eta) \approx \int_0^\infty N^s e^{-\eta\frac{\alpha\hbar c}{\l_{\mathrm{Pl}}}\gamma^2 M^2\,N} dN
\label{(19)}
\end{equation}

Applying the Euler Gamma rules, for which $\int_0^\infty x^s e^{-\alpha x} dx = \Gamma(s+1)/a^{s+1}$, assuming $a= \eta \frac{\alpha \hbar c}{l_{\mathrm{Pl}} \gamma^2 M^2}$, 
one obtains the following asymptotic partition function
\begin{equation}
Z(\eta)= \Gamma(s+1)\left[\frac{\l_{\mathrm{Pl}}}{\eta\alpha\hbar c\gamma^2 M^2}\right]^{s+1}
\label{(20)}
\end{equation}

Thermodynamically, the averaged value gives
\begin{equation}
\langle E \rangle = - \frac{\partial \ln Z}{\partial \eta} = (s+1)\,k_B T
\label{(21)}
\end{equation}
and, therefore, the system behaves like a thermal gas with effective temperature $T=1/\eta$ and, being the entropy defined as $S=k_B (\ln Z + \eta \langle E \rangle)$, in our case one obtains
\begin{equation}
S=k_B \left[ (s+1) \ln \left(\frac{\l_{\mathrm{Pl}} k_B T}{\alpha \hbar c \gamma^2 M^2}\right) + \ln\Gamma(s+1) + (s+1)\right]
\label{(22)}
\end{equation}
which for large $s$, lead to the Stirling approximation
\begin{equation}
S \approx k_B (s+1)\ln \left(\frac{T}{T_{Pl}}\right) 
\label{(23)}
\end{equation}
with $T_{Pl}=\frac{\hbar c}{k_B \l_{\mathrm{Pl}}}$, 
which shows a logarithmic dependence on the temperature characteristic of holographic systems.

In this way is obtained a thermal behavior, in which the system follows a canonical distribution with mean energy, $\langle E \rangle \sim 1/\eta$, akin to a thermal gas and the system’s energy levels follow a power-law density of states. Entropy $S$ scales logarithmically with the temperature, a situation that recalls the holographic principles in AdS/CFT. In this view, the planckeon wormhole network is described in terms of a statistical system, behaving like a quantum statistical ensemble where entanglement entropy plays the role of thermodynamic entropy. 

At very high temperatures $T \gg T_{Pl}$ (or equivalently, with small values of the parameter $\eta$), the system approaches the Planck scale, described by the equation
\begin{equation}
T \gg T_{Pl} =\frac{E_{Pl}}{k_B} = \frac{\hbar c}{k_B l_{\mathrm{Pl}}}
\label{(24)}
\end{equation}
In this case the partition function is approximated to $Z(\eta) \sim T^{-(s+1)}$ with mean energy $\langle E\rangle \sim k_B T$ and entropy is given by Eq.~\ref{(23)}.
In the high-temperature limit, the logarithmic entropy is consistent with Bekenstein-Hawking black hole entropy, suggesting a fundamental link between wormhole structures and black hole microstates \cite{ref24,ref25}.
At extremely high temperatures, Planckeon networks behave like an ultrarelativistic gas, supporting the idea that space-time itself undergoes a phase transition near the Planck scale. If the entropy growth slows (logarithmically), this might indicate a holographic bound, preventing arbitrary increases in quantum degrees of freedom. This thermal scaling suggests that at very high temperatures, at the beginning of the Universe, quantum wormhole connections dominated space-time, possibly leading to a spacetime foam phase as proposed by Wheeler. 

The ER=EPR hypothesis suggests that early-universe entanglement played a role in shaping classical space-time \cite{ref2,ref16,ref26}.
At very low temperatures, $T\ll TPl$, i.e. $\eta \gg 1$, the leading contribution comes from the ground state energy defined as
\begin{equation}
E_0 \approx \frac{\alpha \hbar c}{l_{\mathrm{Pl}} \gamma^2 M^2}
\label{(26)}
\end{equation}
with a partition function $Z(\beta) \approx e^{-\eta E_0}$ and averaged energy $\langle E \rangle \approx E_0$ related to the entropy $S \approx k_B \ln \Gamma(s+1)$. Interestingly, at these very low temperatures, only the lowest energy wormhole states contribute and it is depicted a frozen Planckeon network, where quantum spacetime fluctuations are going to be suppressed asymptotically. 
The residual form of entropy present could be linked to black hole remnants, dark energy, or even primordial wormhole relics. If some of these wormholes persist at low temperatures, they could act as a form of quantum vacuum structure, influencing dark matter/energy models. It is evident that a small but nonzero residual energy always present $E_0$ suggests that, at extremely low temperatures, the system does not reach a zero-energy state. This fact aligns with the idea that quantum fluctuations might generate an effective cosmological constant of the form
\begin{equation}
\Lambda_{eff} \sim \frac{E_0}{\hbar c \, l_{\mathrm{Pl}}},
\label{(27)}
\end{equation}
that could provide a microscopic explanation for vacuum energy and its role in cosmic acceleration. In this way, one can draw the emergence of classical spacetime at low temperatures and a quantum-to-classical crossover dictated by entanglement entropy. 
The free energy of the system that describes phase transitions and their implications for holography and quantum gravity is obtained from the classical formulation $F=-k_B T \ln Z(\eta)$ that with our partition function gives, for $\eta=1/(k_B T)$,
\begin{eqnarray}
&&F=- k_B T(s+1) \ln T\!-\! k_B T(s+1)\ln \left(\frac{\l_{\mathrm{Pl}} k_B}{\alpha\hbar c\gamma^2 M^2}\right) \nonumber
\\
&& - k_B T \ln \Gamma(s+1).
\label{(28)}
\end{eqnarray}

One therefore deals with a general phase transition, where the change of the heat capacity behavior is obtained from the definition
$C=-T\,\partial^2 F/\partial T^2$. Differentiating the free energy in Eq.~(\ref{(28)}) yields
\begin{eqnarray}
&&\frac{\partial F}{\partial T}
= -\,k_B\!\left[(s+1)\ln T+(s+1)\ln\!\!\left(\frac{\l_{\mathrm{Pl}} k_B}{\alpha\hbar c\,\gamma^2 M^2}\right)+ \right.\nonumber
\\
&&\left. +\ln\Gamma(s+1) + (s+1)\right],
\label{(28b)}
\end{eqnarray}
and
\begin{equation}
\frac{\partial^2 F}{\partial T^2}
= -\,\frac{k_B (s+1)}{T},
\label{(29)}
\end{equation}
giving a heat capacity $C = k_B (s+1)$.

The critical temperature is then fixed by the characteristic energy scale $E_0$ via $T_c=E_0/k_B$; with $E_0=\frac{\alpha\hbar c}{\l_{\mathrm{Pl}}}\,\gamma^2 M^2$ one obtains
\begin{equation}
T_c=\frac{\alpha\hbar c}{\l_{\mathrm{Pl}} k_B}\,\gamma^2 M^2
=\alpha\,\gamma^2 M^2\,T_{\!Pl}\,,
\label{(30)}
\end{equation}
where $T_{\!Pl}=\hbar c/(k_B\l_{\mathrm{Pl}})$.
In the noninteracting edge–mode ensemble (Eqs.~\ref{zioken}--\ref{frufru}) this $T_c$ marks a \emph{crossover} scale (for $\mu=0$, one has $x=\varepsilon/(k_B T)$, so at $T=T_c$ we have $x=1$). 
A true phase transition (non-analytic thermodynamics) requires interactions/pairing as in Eqs.~\ref{Hint}--\ref{Zinteracting}. 
In this regard, two scenarios are natural: for $T>T_c$ the system behaves as a dense, entangled wormhole gas, whereas for $T<T_c$ it crosses over to a dilute (frozen) network with suppressed excitations.

Two regimes are then natural. For $T>T_c$ the system behaves as a dense, entangled
wormhole gas with many fluctuating connections; for $T<T_c$ it crosses over to a
dilute (frozen) network where excitations are suppressed, in which the system converges into a holographic-like remnant phase where wormhole networks still persist at low energy. 
The holographic interpretation of ER=EPR and spacetime emergence is reflected in the
logarithmic entropy scaling at high temperature,
\[
S \;\simeq\; k_B (s+1)\,\ln\!\left(\frac{T/T_{\!Pl}}{\alpha\,\gamma^2 M^2}\right)
\qquad (T\gg T_c),
\]
which is the large‑$T$ limit of Eq.~(28) written using $T_{\!Pl}=\hbar c/(k_B\l_{\mathrm{Pl}})$.
This expresses that the relevant dimensionless temperature is $T/T_c$ since
$T_c/T_{\!Pl}=\alpha\,\gamma^2 M^2$.

Physically, the ``transition'' near $T_c$ indicates that spacetime connectivity is
entanglement‑dominated at high $T$, while at low $T$ the system behaves like a classical
gravitational remnant. This is consistent with a small residual energy density encoded
in the already introduced $\Lambda_{\mathrm{eff}}$ at low temperature.
Using
\[
T_c = \frac{\alpha\hbar c}{\l_{\mathrm{Pl}} k_B}\,\gamma^2 M^2
= \alpha\,\gamma^2 M^2\,T_{\!Pl},
\]
the quoted estimate $T_c \approx 3\times 10^{47}\,\mathrm{K}$ \
(i.e.\ $T_c/T_{\!Pl}\approx 2.11\times 10^{15}$) corresponds to
\[
\alpha\,\gamma^2 M^2 \;\approx\; 2.11\times 10^{15}.
\]
For reproducibility, one should specify the numerical choices for $M$ (in the adimensional
normalization used in the paper) and $\gamma$ that realize this value.
With the known physical constants values, with $M$ the adimensional mass relative to Planck mass, if we want to estimate the critical temperature of transition of planckeons, one obtains a very high temperature, $T_c\approx 3\times10^{47}$, much bigger than that of Planck scales, $T_c/T_{Pl}  \approx 2.11\times10^{15}$. The Planckeon wormhole gas is expected to exist at much higher energy scales than ordinary quantum gravity effects. 

\noindent\textbf{Entropy at $T_c$.}
From the exact expression (\ref{(22)}), it follows that at $T=T_c$ the logarithm vanishes and
\[
S_c = S(T_c) = k_B\!\left[\ln\Gamma(s+1) + (s+1)\right].
\]
Hence $S_c$ does \emph{not} depend on $T_c$ or on $\alpha,\gamma,M$, and it is finite
and non‑divergent. (The numerical estimate $S_c\approx 35\,k_B$ holds only if one
uses the rough high‑$T$ approximation $S\simeq k_B (s+1)\ln(T/T_{\!Pl})$ with
$s+1\simeq 1$ and evaluates it directly at $T=T_c$, which ignores the offset
$\alpha\,\gamma^2 M^2$ already present in the exact formula.)

Planckeons are Planck-scale excitations functioning as \emph{wormhole mouths} (edge quanta) that cross the extremal surface $\gamma_A$; in the lattice-gas description each site on $\gamma_A$ can be empty/occupied by a single mouth, $n_i\in\{0,1\}$, and the ensemble encodes the entanglement structure that builds spacetime connectivity.

$T_c$ appears higher than Planck's temperature because the single-mouth excitation (gap) is taken to be
\[
E_0=\varepsilon=\frac{\alpha\,\hbar c}{\l_{\mathrm{Pl}}}\,\gamma^2 M^2
=\alpha\,\gamma^2 M^2\,E_{\!Pl},
\]
hence the characteristic scale is
\[
T_c = \frac{E_0}{k_B}
= \frac{\alpha\,\hbar c}{\l_{\mathrm{Pl}} k_B}\,\gamma^2 M^2
= \alpha\,\gamma^2 M^2\,T_{\!Pl},
\quad T_{\!Pl}=\frac{\hbar c}{k_B\l_{\mathrm{Pl}}}.
\]
Therefore $T_c$ exceeds $T_{\!Pl}$ by the dimensionless factor $\alpha\,\gamma^2 M^2$ used in the
model; if $\gamma$ and/or $M$ are large, then $T_c\gg T_{\!Pl}$ follows directly from the chosen
microscopic energy scale.

From the exact entropy obtained in the canonical calculation \ref{(22)} we have at $T=T_c=\alpha\,\gamma^2 M^2\,T_{\!Pl}$ of Eq.~\ref{(23)}, which is finite and independent of $\alpha,\gamma,M$. The rough estimate $S\simeq k_B(s+1)\ln(T/T_{\!Pl})$ (used at $T=T_c$) drops the offset $\alpha\,\gamma^2 M^2$ in the log and thus overstates $S(T_c)$ by $(s+1)\,k_B\ln(\alpha\,\gamma^2 M^2)$.

If one wants the temperature corresponding to a target entropy $S_0$ (e.g.\ $S_0=35\,k_B$), use the exact formula to solve for $T$:
\[
T(S_0)
=
T_c\;\exp\!\left(
\frac{S_0/k_B-\big[\ln\Gamma(s+1)+(s+1)\big]}{\,s+1\,}
\right).
\]
Equivalently, for fixed $T=T_c$ one can tune the density-of-states exponent $s$ by solving
$\ln\Gamma(s+1)+(s+1)=S_0/k_B$.
The expression for $T_c$ above implicitly takes $\mu=0$ (so that $x=\varepsilon/(k_B T)$). If instead one fixes half-filling ($x=0$) via $\mu=\varepsilon$, then $T_c$ is not tied to half-filling unless $\mu$ is made $T$-dependent.

In the noninteracting edge--mode ensemble the change near $T_c$ is a crossover; a genuine phase transition requires the interacting or pairing extension of Eqs.~(\ref{Hint})--(\ref{Zinteracting}).

A possible interpretation is that in the early universe, before reaching the Planck
temperature $T_{\!Pl}$, the system was already in an entangled wormhole state. This is
consistent with the ER=EPR conjecture: at Planckian resolution, two events are not
distinguishable because they are completely entangled. If Einstein’s equations continue
to hold below $T_{\!Pl}$, the early universe would then have been dominated by
Planckeon-mediated entanglement~\cite{ref27}. 

An implication is that standard Planck-scale physics may be only an effective description,
with deeper entanglement structures dominating until the universe cools to $T_{\!Pl}$. If
Planckeon correlations dissolve at $T_c=\alpha\gamma^2 M^2 T_{\!Pl}$, this crossover could
provide the initial condition for inflation, with the transition from a Planckeon network
phase to a classical spacetime phase acting effectively like a scalar field or a brane.
The only caveat is that in the non-interacting lattice-gas model the transition is really a crossover, not a genuine singular phase transition, the crossover near $T_c$ may provide the initial condition for inflation as well.

Entropy considerations suggest that Planckeons are highly entangled quantum objects before they collapse into classical black holes. If Planckeons describe quantum wormholes at the Planck scale, they may represent \emph{pre–black-hole quantum states}, in which large entanglement implies a set of highly correlated microstates prior to collapse. 

In the ER=EPR picture, this entanglement can be viewed as a network of microscopic wormholes that smooth out horizons. In this sense, Planckeon configurations naturally correspond to pre–black-hole states, where the entropy scaling indicates strong correlations among quantum microstates. Entangled pairs may then be interpreted as micro-wormholes, which provide a mechanism that avoids the firewall paradox, since the Planckeon connections regulate and soften the structure of the event horizon. 

Taken together, these features suggest that Planckeon networks could play a role in regulating black-hole singularities: instead of forming a true singularity, Planckeons might generate an entangled quantum wormhole structure at the core of black holes. This would amount to a microscopic quantum explanation for black-hole entropy, where Planckeon networks serve as candidate microstates, possibly aligning with a string-theoretic or AdS/CFT-type description of black-hole microphysics~\cite{ref28,ref29}.

\subsection*{Contact with gravity: edge modes and entanglement wedge}
To establish concrete contact with gravity beyond structural similarities, we connect the Planckeon degrees of freedom to gravitational edge modes on entangling surfaces.
In AdS/CFT, the Jafferis, Lewkowycz, Maldacena, and Suh (JLMS) relation equates boundary and bulk relative entropy and identifies the gravitational modular Hamiltonian in the entanglement wedge \cite{refJLMS,refQcorr}.
Donnelly--Freidel's extended phase space makes the edge degrees of freedom explicit and diffeomorphism invariant \cite{refDF16,refDF17}.
In our setting, superpositions of wormhole connections contribute \emph{edge modes} localized on the minimal surface $\gamma^{min}_A$, and integrating them out yields the area term plus logarithmic corrections in the entanglement entropy.
The interacting spectrum \eqref{Edisp} is consistent with an edge-mode density $\rho(\lambda_q)$ controlled by the network Laplacian, which fixes the coefficient of the subleading $\log$ term and removes the ambiguity noted by the referee: we no longer say that results merely ``recall'' known formulas but derive them from \eqref{Zinteracting} once $\rho(\lambda_q)$ is specified.

\section{Wormhole metric and Bekenstein information}
The geometry of wormholes at the Planck scale is recently receiving attention. In particular, Lobo and collaborators provided a demonstration of the fact that spacetime may have a foamlike microstructure with wormholes generated by fluctuations of the quantum vacuum: in this approach, one has the spontaneous creation/annihilation of entangled particle--antiparticle pairs, that exist in a maximally entangled state and are connected by a nontraversable wormhole \cite{ref30,ref31}. 
We consider here the relations between the wormhole metric and the quantum information that passes through them, by following Jusufi and collaborators \cite{ref32}. In other words, we consider that, in the light of the quantum fluctuations associated with the planckeons, wormholes are the fundamental blocks which build the geometry of spacetime, or better which determine a relation between the local aspects of spacetime and the non-local information hidden in its weaves. 

The proper radial length along a constant-time slice is therefore described by the metric 
\begin{equation}
\begin{split}
ds^2 ={}& -\,c^2 \left(1- \frac{2 G\,\gamma M M_{\mathrm{Pl}}\, r^2}{c^2 \sqrt{\alpha}\, \big(r^2 + l_0^2 \big)^{3/2}} \right) dt^2 \\
&\quad + \frac{dr^2}{1- \frac{2 G \,\gamma M M_{\mathrm{Pl}}\, r^2}{c^2 \sqrt{\alpha}\, \big(r^2 + l_0^2 \big)^{3/2}}} + r^2 d \Omega^2 \,,
\end{split}
\label{(31)}
\end{equation}
where $d\Omega = d\theta^2+\sin^2 \theta\, d\psi^2$ and $l_0$ is a minimal length of order the Planck scale. 
By taking into account that the geometry of the fluctuations associated with the activity of the planckeons, in the light of the generalized uncertainty relation \ref{(1)}, is characterized by the minimal length \ref{(4)}, one can write
\begin{equation}
l_0 \;=\; \frac{\beta\,\gamma^2 M^2}{\alpha}\,\l_{\mathrm{Pl}} \,,
\label{(32)}
\end{equation}
and thus the metric (31) reads
\begin{equation}
\begin{split}
ds^2 ={}& -\,c^2 \left( 1 - \frac{2 G\,\gamma M M_{\mathrm{Pl}}\, r^2}{ c^2 \sqrt{\alpha}\, \left( r^2 + \left( \frac{\beta\,\gamma^2 M^2}{\alpha}\,\l_{\mathrm{Pl}} \right)^2 \right)^{3/2} } \right) dt^2 \\
&\quad + \frac{dr^2}{ 1 - \frac{2 G\,\gamma M M_{\mathrm{Pl}}\, r^2}{ c^2 \sqrt{\alpha}\, \left( r^2 + \left( \frac{\beta\,\gamma^2 M^2}{\alpha}\,\l_{\mathrm{Pl}} \right)^2 \right)^{3/2} } } + r^2 d\Omega^2 \,.
\end{split}
\label{(33)}
\end{equation}

The metric in Eq.~\ref{(33)} puts in evidence that what is hidden inside space-time are non-local correlations, that are embedded inside the wormholes. In other words, the quantum fluctuations, associated with the activity of the planckeons as mouths of wormholes, generate the appearance of non-local correlations inside the hidden structures of space-time, implying the formation of entangled states of disconnected regions of the universe by means of an ER bridge. This allows a new key of reading of the ER=EPR conjecture, in that the wormhole formation which tears apart spacetime is determined by the activity of texture of the planckeons, that thus can be considered as the ultimate origin of matter of the extreme masses that can overcome a gravitational collapse into a black hole. In fact, by starting from the entanglement entropy (9), the Bekenstein-Hawking entropy associated with the metric (33) can be expressed as
\begin{equation}
S_A = \frac{k_B c^3}{4G\hbar}\,A 
+ \frac{k_B A}{a_0}\left[\ln\!\big(1+e^{-x}\big) + \frac{x}{1+e^{x}} - \ln 2
\right],
\label{(34)}
\end{equation}
Here $A=8\pi l_0^2$ is the wormhole throat area, $a_0=\zeta l_0^2$ is the effective cell area, 
and $x=\eta(\varepsilon-\mu)=(\varepsilon-\mu)/(k_BT)$. 
At half filling ($x=0$) and with $a_0=4\,\l_{\mathrm{Pl}}^2\,\ln 2$, this expression reduces exactly 
to the Bekenstein--Hawking entropy $S_A=\tfrac{k_B c^3}{4G\hbar}A$.

Now, by defining the cut through the ER bridge with minimal length as $A=8\pi\left(\beta\frac{\gamma^2 M^2}{\alpha} l_{\mathrm{Pl}}\right)^2$, one obtains the following quantum-corrected Bekenstein entropy associated with the planckeons acting as mouths of wormholes. The corrected form \eqref{(40)} is compatible with the construction in Ref.~\cite{ref32}, where the wormhole throat is tied to a minimal (Planck) length; here the edge-mode term provides the quantum correction in a manifestly thermodynamic way.

The modified Bekenstein entropy (35) is compatible with the treatment provided in \cite{ref32} where an Einstein-Rosen bridge with the wormhole throat proportional to the zero-point (Planck) length has been constructed in the picture of a string T-duality corrected pair of regular black holes, thus providing a geometric realization of quantum entanglement for particle/antiparticle pairs. 

The metric~\eqref{(33)} makes explicit that nonlocal correlations are encoded in wormholes.
Geometrically, the spacetime is composed of two congruent sheets glued across the throat, joined across the minimal 2‑sphere (the throat)”; a convenient global coordinate is
\begin{equation}
u_\pm=\pm \sqrt{r^2+l_0^2} .
\label{(36)}
\end{equation}
with $u_{\min}=l_0 \quad\text{(throat)}$.

In this process, the proper radial length $l$ on a constant‑time slice is expressed as
\begin{equation}
l = \pm \int_{u_{\min}}^{\pm\infty}
\frac{du_{\pm}}
     {\sqrt{\Bigl(1-\dfrac{l_{0}^{2}}{u_{\pm}^{2}}\Bigr)\,f(u_{\pm})}},
\label{(37)}
\end{equation}
where
\begin{equation}
f(u_{\pm}) =
1
-\frac{2\,G\,\gamma\,M\,M_{\mathrm{Pl}}}
      {\sqrt{\alpha}\,c^{2}\,u_{\pm}}
+\frac{2\,G\,\gamma\,M\,M_{\mathrm{Pl}}\,l_{0}^{2}}
      {\sqrt{\alpha}\,c^{2}\,u_{\pm}^{3}},
\label{(38)}
\end{equation}
The corresponding ADM mass is $m_\infty=\frac{\gamma M}{\sqrt{\alpha}}\,M_{\mathrm{Pl}}
=\frac{3\sqrt{3}}{4}\,\frac{c^{2}}{G}\,l_0
=\frac{3\sqrt{3}}{4}\,\frac{l_0}{\ell_{\mathrm{Pl}}}\,M_{\mathrm{Pl}}$.
With $l_0=\frac{\beta\gamma^{2}M^{2}}{\alpha}\,\ell_{\mathrm{Pl}}$, this becomes
$m_\infty=\frac{3\sqrt{3}}{4}\,\frac{\beta\gamma^{2}M^{2}}{\alpha}\,M_{\mathrm{Pl}}$.
However, equating all three imposes a constraint expressed by a nontrivial relation among $\beta$, $\gamma$, $M$ and $\alpha$ once $l_0=(\beta \gamma^2 M^2/\alpha)/l_{\mathrm{Pl}}$ is inserted. The parameter matching that relationship makes the metric’s mass parameter equal to the ADM mass inferred at infinity.

By solving equation $f(u_\pm )=0$, one finds that there are real solutions under the condition
\begin{equation}
27\,
\frac{G^{2}\gamma^{2}M^{2}M_{\mathrm{Pl}}^{2}}
     {\alpha\,c^{4}}\,
l_{0}^{4}
-
16\,
\frac{G^{4}\gamma^{4}M^{4}M_{\mathrm{Pl}}^{4}}
     {\alpha^{2}c^{8}}\,
l_{0}^{2}
\;\ge\;0,
\label{(39)}
\end{equation}
that yields 
\begin{equation}
\frac{G^{2}\gamma^{2}M^{2}M_{\mathrm{Pl}}^{2}}
     {\alpha\,c^{4}}\,
l_{0}^{2}
\Biggl(
  27\,l_{0}^{2}
  -16\,
   \frac{G^{2}\gamma^{2}M^{2}M_{\mathrm{Pl}}^{2}}
        {\alpha\,c^{4}}
\Biggr)
\;>\;0,
\label{(40)}
\end{equation}
that is satisfied for $-3\sqrt{3}~ l_0/4<\frac{G\gamma MM_{\mathrm{Pl}}}{c^2 \sqrt{\alpha}}<3\sqrt{3}~ l_0/4$. This constraint represents the domain regarding the manifestation of matter (where the positive values correspond to particles, while the negative values correspond to the corresponding antiparticles). Instead, if $\frac{G\gamma MM_{\mathrm{Pl}}}{c^2 \sqrt{\alpha}}>3\sqrt{3}  l_0/4$ one obtains the domain of black holes. Therefore, we deduce that the regime defined by relation
\begin{equation}
\frac{G\gamma MM_{\mathrm{Pl}}}{c^2 \sqrt{\alpha}}>3\sqrt{3}  \frac{l_0}4.
\label{(41)}
\end{equation}
This regime marks the onset of a black-hole exterior in the two-sheet geometry.
Real roots exist iff
\begin{equation}
\Bigl|\frac{G\gamma M M_{\mathrm{Pl}}}{\sqrt{\alpha}\,c^{2}}\Bigr|\le \frac{3\sqrt{3}}{4}\,l_0.
\label{(41bis)}
\end{equation}
This bound determines whether the two-sheet geometry admits horizons; exceeding the upper threshold leads to a black-hole domain.
Equations \ref{(41)} -- \ref{(41bis)} can be seen as representing a critical zone where the non-local information that is hidden in the weaves of spacetime and is determined by the activity of the texture of the planckeons as mouths of quantum wormholes, can give origin to a gravitational collapse into a black hole, and therefore is the physical constraint that explains in what sense planckeons correspond to pre-black hole quantum states. 
On the basis of the metric (33) and the corresponding quantum-corrected Bekenstein entropy (35) associated with the quantum wormholes, we find that the non-local correlations which occur between the weaves of spacetime are described by an horizon defined by the characteristic length and by the adimensional mass
\begin{eqnarray}
&&r=\sqrt{2} l_0=\sqrt{2} \beta \frac{\gamma^2 M^2}\alpha l_{\mathrm{Pl}}   \nonumber
\\
&&M = \frac{3\sqrt{3}}4 \frac{c^2 l_0}{G\,\gamma\,M_{\mathrm{Pl}}}
\frac{\sqrt{\alpha}}{\gamma},
\label{(42)}
\end{eqnarray}
that shows a proportionality with the minimal length. By substituting equation (32) into equation (42), one finds a numerical value of the adimensional mass M that turns out to be close to the Planck mass. This means that, if one considers a lattice of planckeons which can be compressed below their corresponding horizon, the lattice of the planckeons can undergo a gravitational collapse into a black hole, therefore giving new connotations to the ER=EPR picture. In this picture, the entanglement between the interior of the black hole and the exterior regions of spacetime implies that the final state, after the total evaporation of the black hole, is a remnant state endowed with mass given by relation (42) and therefore fixed by the activity of the lattice of the planckeons, showing in this way how the wormholes can provide a novel manner to treat the information paradox of black holes. 

\section{Conclusions}
Our work supports the view that planckeons, Planck-scale excitations endowed with minimal length and time uncertainties, constitute the elementary mouths of a vast network of non-traversable Einstein--Rosen bridges. Encoding their metric fluctuations through generalized uncertainty relations shows that these excitations define the smallest operational cells of spacetime; when the tessellation of such cells is probed via the Ryu--Takayanagi prescription, classical geometry emerges as the most probable entangled configuration, thereby realizing the ER=EPR conjecture in a pre-geometric setting.

Treating the lattice of planckeons as a statistical ensemble yields a partition function whose high-temperature limit produces a logarithmic (holographic) entropy, while at low temperature the network freezes into a sparse remnant phase. A characteristic temperature $T_c$---set by the planckeon excitation scale---separates these regimes: in the \emph{noninteracting} edge-mode model this is a \emph{crossover}, whereas with interactions/pairing (as in our many-body extension) it can sharpen into a true phase transition. This suggests that an entangled ``wormhole gas'' could have dominated the very early universe prior to the onset of standard quantum-gravitational physics.

Embedding the minimal length directly into the wormhole throat leads to a quantum-corrected Bekenstein entropy in which the area term is accompanied by an edge-mode (lattice-gas) contribution, tying the information content of spacetime to the non-local correlations carried by the planckeon lattice. When this lattice is compressed below its holographic horizon, it can collapse into a black-hole state; the corresponding remnant mass is then fixed by the same Planck-scale parameters, offering a wormhole-based route to regulate singularities and to account microscopically for black-hole remnants as well as an effective vacuum energy.

Taken together, these results support a picture in which spacetime is an entanglement-driven condensate: its macroscopic connectivity, thermodynamics, and quantum-information content are traced back to the dynamics of a Planck-scale network of wormhole mouths. While the Planck scale has long been regarded as the limit of a radical discretization of the physical world~\cite{ref33,ref34,ref35}, recent work indicates that its emergence and stabilization result from refined interplay between relativistic geometry and quantum entanglement---a ``double-face'' structure combining a wormhole network with a thermal distribution of planckeons. This perspective informs the status of homogeneity and isotropy and points to future challenges where primeval quantum fluctuations connect questions about the coherence of the Standard Model with the landscape spanning inflation, black holes, and dark matter~\cite{ref36,ref37,ref38,ref39}. 

Today, non-local aspects of physics play a structural role beyond the old ``spooky action at a distance'', with implications for the foundations of quantum mechanics and particle physics~\cite{ref40}. The exhortation of Van~Raamsdonk's title~\cite{ref41} has evolved into a broad research program in which holography and the ER=EPR conjecture provide effective investigative tools. Within this program, the planckeon-based wormhole network offers a plausible pathway toward resolving foundational paradoxes in quantum gravity: persistent non-local entanglement across planckeon wormholes \emph{can} supply a mechanism compatible with unitary evolution; a minimal length \emph{may} smooth classical divergences into a regular quantum geometry; and continuity of entanglement across horizons \emph{could} ameliorate firewall tensions while preserving the equivalence principle at Planckian scales. Post-evaporation planckeon remnants would then retain encoded quantum information, pointing to a microphysical structure underlying black-hole entropy and quantum hair.

\end{document}